\documentclass[aps,prd,showpacs,twocolumn]{revtex4}

\usepackage{latexsym}
\usepackage{amsfonts,amssymb}
\usepackage{graphicx,epsfig}
\usepackage{psfrag}

\newcommand{\be}{\begin{equation}}
\newcommand{\ee}{\end{equation}}
\def\bq{\begin{eqnarray}}
\def\eq{\end{eqnarray}}
\def\n{\nonumber}

\def\s{\sigma}

\def\Om{\Omega_{\rm m}}
\def\la{\lambda}
\def\tla{\Lambda}

\begin{document}

\title{Brane curvature and supernovae Ia observations}

\author{Parampreet Singh\footnote{e-mail: param@iucaa.ernet.in}, R. G. Vishwakarma\footnote{e-mail: vishwa@iucaa.ernet.in},  Naresh
Dadhich\footnote{e-mail: nkd@iucaa.ernet.in }}
\address{Inter-University Centre for Astronomy and Astrophysics,\\
Post Bag 4, Ganeshkhind, Pune-411 007, INDIA.}

\begin{abstract}
It is well known that modifications to the Friedmann equation on a warped
brane in an anti de Sitter bulk  do not provide any low energy distinguishing
feature from standard cosmology. However, addition of a brane curvature
scalar in the action  produces effects which can serve as a distinctive
feature of brane world scenarios and can be tested with
observations. The fitting of such a model with supernovae Ia data (including SN 1997ff at
$z\approx1.7$) comes out very well and predicts an accelerating universe.
\end{abstract}

\pacs{04.50.+h, 98.80.-k, 97.60.Bw }

\maketitle

In recent times there has been considerable activity in
large extra dimensions and the recovery of the four dimensional General
Relativity (GR) as an effective theory from a more fundamental theory.
The basic idea in these theories is the existence of a higher dimensional
bulk in which our universe, called 3-brane is sitting as a hypersurface. The
standard model matter fields are confined to the brane while gravity can, by
its universal character, propagate in all (including extra) dimensions.
The large extra dimensions also promise to solve the mass
hierarchy problem of the standard model of particle physics.
A particular example is the model of Randall and Sundrum \cite{rs}, which
has a warped extra dimension and has attracted a lot of interest.

In this model the authors considered a 5-D anti de Sitter
(AdS) bulk with a $Z_2$-symmetric extra dimension and by a proper fine
tuning of bulk parameters they obtained a flat
static brane with a vanishing effective cosmological constant on the brane.
The standard Newtonian potential is recovered in the model
with $r^{-3}$ corrections arising from the massive Kaluza-Klein (KK) modes
on the brane. The $Z_2$-symmetry is motivated by the reduction of M theory
to $E_8 \times E_8$ heterotic string theory \cite{hw}.

There have been
many generalizations of this model and various cosmological
issues have been addressed therein ( see for example \cite{cosmo}).
The recovery of Newtonian force law in this kind of models is a non-trivial
task and requires solving the perturbations
of the bulk metric. It is not clear a priori whether Newtonian gravity can
always be recovered in such models. A counter example is provided by the
conformally non-flat Nariai metric, for which there exists no massless
graviton on the brane \cite{sd1}. On the other hand, this recovery has been
established for the Schwarzschild-AdS (S-AdS) bulk (which is also conformally
non-flat) with an FRW brane metric and it gives rise to various
possible physical universes
including the one with effective positive cosmological constant \cite{sd2}.
 Hence it could be envisioned that our FRW universe
may be expanding out of a Schwarzschild black hole in an AdS bulk space.

Using the Israel junction conditions \cite{israel}, one can write the
analogue of the Friedmann equation
on the brane  \cite{muk}
\be
H^2 = \frac{8 \pi G}{3} \, \rho \, \left(1 + \frac{\rho}{2 \s} \right) +
\frac{\la}{3} + \frac{C}{S^4} - \frac{k}{S^2} \label{eq:fosb},
\ee
where $G$ and $\la$ are respectively 4-D gravitational and
cosmological constants, $\rho$ is the matter density on the brane,
$S$ is the scale factor,
$\s$ is the brane tension  and  $C$ is the mass parameter of the bulk black
hole. The  four dimensional effective constants are related to the
five dimensional bulk constants through the Israel junction conditions,
\be
\lambda = \frac{\lambda_5}{2} + \frac{16 \pi^2}{3} \, G_5^2 \sigma^2,
\, ~ ~ G = \frac{4 \pi}{3} \, G_5^2 \, \s
\ee
where $\lambda_5 = - 6/l^2$ and $G_5$ are the 5-D cosmological and
gravitational constants. Here $l$ is the radius of curvature of the bulk
spacetime.

The model differs from the standard cosmology in the following two terms
in eq.(\ref{eq:fosb}).
First, the density-squared term representing the local effects, which
arises due to corrections in stress tensor by imposing junction conditions.
This term decays as $1/S^8$ in 
the radiation dominated epoch and  would be insignificant even at the time of 
big bang nucleosynthesis (BBN) \cite{ichiki}.
Second, the term varying as $1/S^4$, commonly known as 
the {\it dark radiation}, represents the non
local bulk effect in terms of mass of the black hole. This term
enters into eq.(\ref{eq:fosb})
through the projection of bulk Weyl curvature on the brane and behaves like
radiation. It can affect the BBN and hence can be
 constrained by observations.
As we shall show in the following, the {\it dark radiation} term
can be safely neglected at the epoch even as remote in the past as
$z\approx 1.7$, which is the highest redshift observed so far in the case of
the supernovae Ia.
Thus there survives no observational effect of these brane world modifications
and the model reduces to the standard FRW model without having any distinguishing
features of its own.

It is interesting to note that if one considers
the brane models with Minkowski bulk
\cite{mbulk1,mbulk2,mbulk3} the deviations from standard cosmology are
expected at a cross over scale $r_c$ defined as the ratio $M_4^2/2 M_5^3$,
where $M_4$ and $M_5$ are respectively the 4-D and 5-D Planck masses. The
action in these models contains a term proportional to the curvature scalar
on the brane which also helps in recovery of 4-D Newtonian gravity on
the brane. The cosmological consequences of such an introduction in
these models were first studied in Ref. \cite{mbulk2,mbulk3} and the
Friedmann equation with the cosmological constant on the brane 
in these models has also been derived \cite{mbulk2,hy}.
The comparison of these models
with cosmological observations has also been pursued \cite{mbulkobs}.

The term proportional to curvature scalar can be viewed as a quantum
correction and is also usually needed for a proper definition of stress-energy
tensor on the boundary of S-AdS spacetime \cite{ac}.
A variant of the AdS bulk brane model was considered by  Kim, Lee and Myung
\cite{myung}
in which they incorporated a brane curvature scalar term in the bulk-brane
action via a small coupling parameter. 
This model provides a departure
from the standard cosmology at late time evolution of the universe. 
Here, our main aim is to test the model against
the supernovae Ia observations and study its cosmological consequences.
In fact, the model fits the SN Ia data very well.

The brane in this model is a hypersurface given by the metric
\be
d s^2 =  - d\,  t^2 +  S^2(t)\,  \left(\frac{d r^2}{1 - k r^2} + r^2d \, \theta^2 + r^2\sin^2 \,
\theta \, d \, \phi^2 \right)
\ee
embedded in a bulk  composed of two patches of negative
$\lambda_5$ spacetime whose action is
\bq
S &=& \n  \frac{1}{16 \pi G_5} \, \int d x^5 \, \sqrt{-g} \, \left( R + \frac{12}{l^2} \right) \\ &+& \n  \frac{1}{8 \pi G_5} \, \int d x^4 \sqrt{-h } K 
+ \s \, \int d^4 x \,\sqrt{-h } \\
&+&  \frac{b}{16 \pi G_5} \int d^4 x \, \sqrt{-h} \, \frac{l}{2} \, R^{(4)},
\eq
where $K$ is the trace of the extrinsic curvature of the brane. The first term in the action is the standard Einstein-Hilbert
action, the second term is the Gibbons-Hawking term which is
necessary for the variational problem to be well defined and the third term is
the contribution of brane tension to the action. We are interested in looking
at the incorporation of the brane curvature scalar term in the action 
 coupling through the parameter $b$ which is small so that its
higher order terms in effective Einstein's equations on the brane 
 can be neglected at the present epoch.
For this model the modified Friedmann equation \cite{myung}, in low
energy limit ($\rho \ll \s$), becomes
\be
H^2 =   \frac{8 \pi G}{3} \, \rho  + \frac{\tla}{3}
+ \frac{\cal C}{S^4}  - \frac{(k - \alpha)}{S^2}\label{eq:nfeq}
\ee
where
\bq
\alpha &=& b \, \bigg[ \sqrt{\frac{\pi \, G \, \s}{3}} \, l - \frac{1}{16 \, \pi \, G \, \s \, l^2}\bigg] ,\\
\tla &=& \la - \frac{3 \, b}{l} \, \sqrt{3  \pi  G  \s} \, \bigg[\frac{8 \, \pi \, G \, \s \, l^2}{9} + \frac{1}{4  \pi  G  \s  l^2} - 1 \bigg] ,\\
{\cal C} &=& C \, \bigg[1 + b \left(\frac{3 - \beta^2}{3 \, \beta} \right)\bigg], ~~~~ \beta = 4 \, l \, \sqrt{\frac{\pi \, G \, \s}{3}}.
\eq
 There would also be higher order terms in $S$ in eq.(\ref{eq:nfeq}),
which have not been included for their effect being insignificant
at the epochs of redshifts upto 1.7.
The term $\alpha/S^2$ would be 
dominant over the density and dark energy terms  at late times of 
expansion of the universe. It may be noted that $1/S^2$ fall off is typical
of $\rho + \Sigma_i \, p_i = 0$ which is a characteristic of topological
defects like cosmic strings and textures \cite{cs1}. The inclusion of
topological defects in standard cosmology has been widely considered (see
for example \cite{cs2}).

The {\it  dark radiation} term in eq.(\ref{eq:nfeq}) now contains additional
contributions from the bulk parameters. In order to calculate the relative
contribution of this term to the expansion dynamics, we write eq.(\ref{eq:nfeq})
in a more convenient form as

\be
H^2(z) = H^2_0 \, \bigg[ \Omega_{{\rm m} 0} \, (1 + z)^3 +  \Omega_{\Lambda 0}+\Omega_{\rm dr 0}(1 + z)^4
-\Omega_{k_{\rm eff}0} \, (1 + z)^2 \bigg], \label{eq:ofe3}
\ee
where $k_{\rm eff} \equiv k - \alpha$ \, and we have defined the various energy density components, in units of the critical
density, in the following form.
\be
 \Om \equiv \frac{8 \pi G}{3 H^2} \rho, \,\hspace{0.25cm}
 \Omega_{\Lambda} \equiv \frac{\tla}{3 H^2}, \, ~~
\Omega_{\rm dr}\equiv \frac{{\cal C}}{S^4 H^2}, ~~
\Omega_{k_{\rm eff}} \equiv \frac{(k - \alpha)}{S^2 H^2}. \label{eq:omegas}
\ee
The subscript 0 denotes the value of the quantity at the present epoch.
The BBN constrains the {\it  dark radiation}
density to be between ($-$1.23 and  0.11) $\times$ $\rho_{\rm r0}$ \cite{ichiki},
where  $\rho_{\rm r0}$ is the present energy density of the photon background with
$\Omega_{\rm r0} \approx 2.5 \times 10^{-5}$.
Thus the upper BBN constraint imply that at the highest SN redshift $z=1.755$, the
contribution of the (positive) {\it  dark radiation} term in eq.(\ref{eq:ofe3}) is only
0.003 \% of the contribution from the $\Omega_{\rm m}$ term for $\Omega_{\rm m0}$
as low as 0.33 (which comes from the various recent measurements of 
$\Omega_{\rm m0}$ as $\Omega_{\rm m0}=0.330\pm 0.035$ at one 
sigma level \cite{turner1}).
Similarly for the lower constraint (i.e., negative {\it dark radiation}),
the corresponding contribution is 0.03 \%.
The {\it dark radiation} term, hence, can be
safely neglected while considering the SN Ia observations and eq.(\ref{eq:ofe3})
reduces to

\be
H(z) = H_0 \, \bigg[ \Omega_{\rm m0} \, (1 + z)^3 +  \Omega_{\Lambda 0} -
\Omega_{k_{\rm eff}0} \, (1 + z)^2 \bigg]^{1/2} \label{eq:ofe2}
\ee
implying that $\Omega_{\rm m0} + \Omega_{\Lambda 0} = 1 + \Omega_{k_{\rm eff} 0}$.
Note that $\alpha$ does not enter into the metric  on the brane
but it enters in the brane Friedmann equation eq.(\ref{eq:nfeq})
through the Israel junction conditions.
This is equivalent to shifting the curvature index $k$ of the standard
cosmology by an amount $-\alpha$. This term ($k-\alpha)/S^2$ can be
neglected in the early universe.
However, it dominates at the present epoch and predicts
significant departures from the standard cosmology,
especially the expansion dynamics of the universe.

In order to test the model with the supernovae Ia observations, we derive,
in the following, the magnitude-redshift relation. We note that
the {\it luminosity distance} $d_{\rm L}$ of a source of redshift $z$,
located at a radial coordinate distance $r$, is given by
$d_{\rm L} = (1 + z) \,S_0 \, r$ where
 $r$ can be calculated from the metric as
\bq
&=& \n \sin^{-1}r, ~~~\mbox{when} ~~~ k = 1 \\
\frac{1}{S_0} \, \int_0^z \, \frac{d z'}{H(z')}&=& \n r, ~~ ~~ ~~ ~~ ~~ ~\mbox{when} ~~~ k = 0 \\
&=& \sinh^{-1}r ~~~\mbox{when} ~~~ k = - 1. \label{eq:rdist}
\eq
The apparent magnitude $m$ of the source is given by
\be
m (z) = {\cal M} + 5 \, \log [{\cal D}_{\rm L}(z) ] \label{eq:mageq}
\ee
where ${\cal M} \equiv M - 5 \,\log H_0 +$ constant, $M$ is the absolute
luminosity of the source
and  ${\cal D}_{\rm L} \equiv H_0 d_{\rm L}$ is the dimensionless luminosity distance.
The present {\it radius} of the universe $S_0$, appearing in
$d_{\rm L}$, can be calculated from eq.(\ref{eq:omegas}) as
\be
S_0^2 H_0^2 =\frac{k-\alpha}{\Omega_{\rm m0}+\Omega_{\Lambda_0}-1} ~ ~ ~ ~ (k \neq \alpha).  \label{eq:szero}
\ee

We can now calculate, by using eqs.(\ref{eq:ofe2}-\ref{eq:szero}),
the predicted value of the apparent magnitude at a given redshift
if we know the values of the parameters $\Omega_{\rm m0}, \Omega_{\Lambda 0}$, $\alpha$
and ${\cal M}$. In order to test the model, we consider the data
on the redshift and magnitude of a sample of 54 supernovae of type Ia
considered by Perlmutter et al (excluding 6 outliers from the full
sample of 60 supernovae) \cite{perl},
together with the recently observed supernova 1997ff at $z=1.755\pm0.05$
(with $m^{\rm corr}=25.68\pm 0.34$) the
highest redshift supernova observed so far \cite{riess}.
The best-fitting parameters are obtained by minimizing $\chi^2$, which is
defined by
\be
\chi^2 = \sum_{i = 1}^{55} \, \Bigg[\frac{m_i^{\rm corr} - m(z_i)}{
\delta m_i^{\rm corr}} \Bigg]^2,
\ee
where $m_i^{\rm corr}$ refers to the effective
magnitude of the $i$th supernovae which has been corrected by the lightcurve
width-luminosity correction, galactic extinction and the K-correction from
the differences of the R- and B-band filters and the dispersion
$\delta m_i^{\rm corr}$ is the uncertainty in $m_i^{\rm corr}$.

\begin{figure}[tbh!]
\epsfig{figure=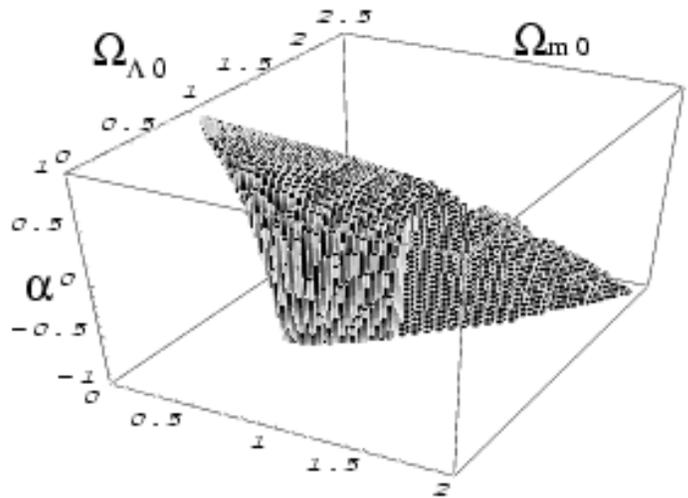,height=2.6in,width=3.6in,angle=0}
{\caption{\small A parameter space of interest at 68 percent confidence
level is shown by the conical volume for $\alpha \geq -1$.}}
\end{figure}

The model fits the data for a wide range of parameters. In Fig. 1,
the conical volume shows the 68 \% confidence region which
 gets contributions only from the closed and flat
models. The open model contributes only at higher confidence levels,
more than 85 \%. In the case of flat ($k = 0$) model (where
$\Omega_{k_{\rm eff} 0}$ is not necessarily zero), $m$, and hence $\chi^2$,
are independent of $\alpha$ which is constrained only through eq.(\ref{eq:szero}).
The best-fitting
parameters are obtained as
$\Omega_{\rm m0} = 0.35$, $\Omega_{\Lambda 0} = 1.12$, $\alpha = 0.69$  and ${\cal M} = 23.91$
with $\chi^2 = 56.85$ at 51 degrees of freedom (dof), which represents a
fit as good as for the best-fitting standard model ($\Omega_{\rm m0} = 0.87$,
$\Omega_{\Lambda 0} = 1.51$ and ${\cal M} = 23.9$ with $\chi^2 = 56.89$ at 53 dof
\cite{vishwa1}).
However, these  models are not consistent with the recent
anisotropy measurements of the cosmic microwave background radiation (CMBR),
which imply that $k\approx 0$ \cite{cmb}.
However, the parameter space is large enough (as in the case of standard 
model), as is clear from Fig 1, to accommodate
with the CMBR predictions.
Additionally, we find that the model also fits the data very well for the case
$k=0$ for a wide range of parameters giving $\Omega_{\rm m0} = 1.35$, $\Omega_{\Lambda 0} = 1.54$,
${\cal M} = 23.93$ with $\chi^2 = 58.8$ (at 52 dof) as the best-fitting solution. Though this best fitting $ \Omega_{\rm m0}$ is higher than the favoured
value  0.33, however the parameter space is sufficiently large to accommodate
low  $ \Omega_{\rm m0}$ values as has been shown in Fig 2, where  
we have shown the 68 \% and 95 \% confidence regions for this case.
Note that, unlike the standard model,  the points in the figure are not
confined only on a straight line since for a given pair $\Omega_{\rm m0}$ and
$\Omega_{\Lambda 0}$, model is degenerate in $\alpha$ and $S_0$.
For the case $k=0$,  $d_{\rm L}$ (hence $m$ and $\chi^2$) is independent 
of $\alpha$.
If we fix  $\Omega_{\rm m0}$ to 0.33, then the best-fitting solution (for $k=0$) is obtained as
$\Omega_{\Lambda 0} = 0.7$, ${\cal M} = 23.96$ with $\chi^2 = 62.0$ at 53 dof.
The allowed ranges of $\Omega_{\Lambda0}$
at 68 \% and 95 \% confidence levels are obtained, respectively, as $0.53-0.84$
and $0.34-0.97$.
It should be noted that for this geometrically flat ($k=0$) case,
the model is not dynamically flat
($\Omega_{\rm total}\equiv \Omega_{\rm m}+\Omega_{\Lambda}\neq 1$) and reduces
to the standard flat model, $\Omega_{\rm total}=1$ only when $\alpha = 0$.

We also notice that the exclusion of SN 1997ff from the sample has only a marginal
effect on our results. For example, the global best-fitting model from the Perlmutter
et al' sample of 54 points yields
$\Omega_{\rm m0} =0.34$, $\Omega_{\Lambda 0} = 1.14$, ${\cal M} = 23.91$, $\alpha=0.7$ with
$\chi^2 = 56.83$ at 50 dof. The best-fitting model for the case $k=0$, from this
sample, yields $\Omega_{\rm m0} = 1.7$, $\Omega_{\Lambda 0} = 1.9$,
${\cal M} = 23.91$ with $\chi^2 = 56.9$ (at 51 dof).

\begin{figure}[tbh!]
\epsfig{figure=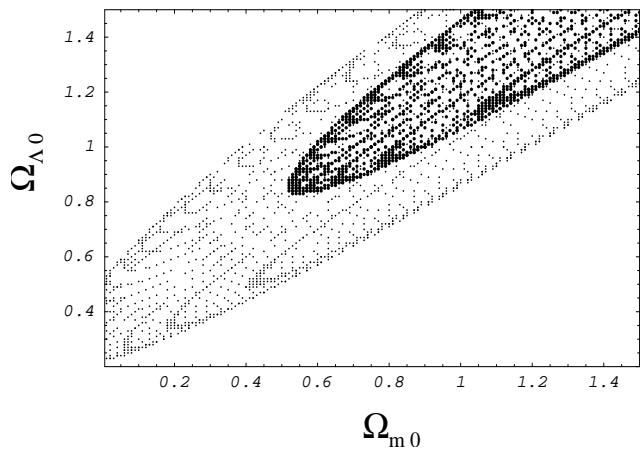,height=2.3in,width=3.3in,angle=0}
{\caption{\small  The elliptic regions at 68 \% confidence level (dark shaded)
and 95 \% confidence level (light shaded) are shown for the geometrically
flat model ($k=0$). }}
\end{figure}

The present value of the Hubble parameter $H_0$ sets the age of the universe.
A large number of independent methods appear to converging on a value
of $H_0$ in the range (60 $-$ 80) km s$^{-1}$ Mpc$^{-1}$ \cite{hvalue},
which sets the age of the best-fitting flat standard model
in the range (11.4 $-$ 15.2) Gyr. If $H_0$ is towards the lower side of this
range, there is no  serious age problem
with the standard cosmology in the view that the age of the globular clusters
$t_{\rm GC} = 12.5 \pm 1.2 \,$ Gyr \cite{gc} and the age of Milky Way
$t_{\rm MW} = 12.5 \pm 3 \,$ Gyr \cite{mw}.
However, if $H_0$ shifts on the higher side (as is claimed by the recent
measurements of $H_0= 72 \pm 7 \,$ km s$^{-1}$ Mpc$^{-1}$ by using Hubble
Space Telescope \cite{turner2}), the standard model might get uncomfortable
with its age.
 However, one can obtain larger age $t_0$
if larger values of $\Omega_{\Lambda0}$ are allowed, as is clear from Fig 3,
where we have plotted ($t_0$) as a function of $\Omega_{\Lambda 0}$ freezing
$\Omega_{\rm m0}$ at its observed value 0.33. The expression for the age of the universe
is the same in both models and is given by
\be
t_0 = H_0^{-1} \int_0^\infty \frac{(1+z)^{-1}\, d z}{\{(z\Omega_{\rm m0}+1)(1+z)^2-
\Omega_{\Lambda 0} z(z+2)\}^{1/2}} .
\ee
In the favoured standard model ($\Omega_{\rm total}=1$), $t_0$ can be increased
only by increasing
$\Omega_{\Lambda 0}$ (i.e. by reducing $\Omega_{\rm m0}$), which
does not seem likely as the recent measurements give very narrow range of
$\Omega_{\rm m0}$, as mentioned earlier. However, there is no such constraint on the
present model and fairly large values of $\Omega_{\Lambda 0}$ are allowed
by the data: $\Omega_{\Lambda 0}$ can rise as high as 0.97 at 2 sigma level
(by fixing $\Omega_{\rm m0}=0.33$),
which can increase the age as large as 14 Gyr even for $H_0$ as high as
72 km s$^{-1}$ Mpc$^{-1}$.

\begin{figure}[tbh!]
\epsfig{figure=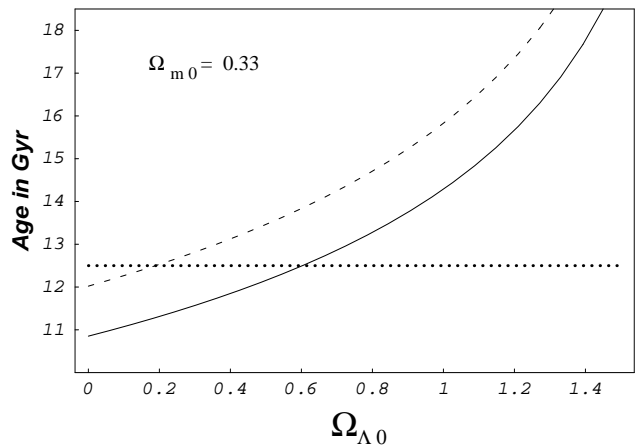,height=2.3in,width=3.3in,angle=0}
{\caption{\small Age of the universe is plotted as a function of
$\Omega_{\Lambda 0}$ for the cases $H_0 = 72$ km s$^{-1}$ Mpc$^{-1}$ (solid curve)
and $H_0 = 65$ km s$^{-1}$ Mpc$^{-1}$ (dashed curve).
The value of $\Omega_{\rm m0}$ has been fixed at its observed value 0.33.
The horizontal dotted line represents the
age of the globular clusters $t_{\rm GC}=12.5$ Gyr.}}
\end{figure}

In Fig 4, we have
shown the actual fitting of the best-fitting geometrically flat model
(fixing $\Omega_{\rm m0}=0.33$) with the data and compared it with the favoured
standard model.

\begin{figure}[tbh!]
\epsfig{figure=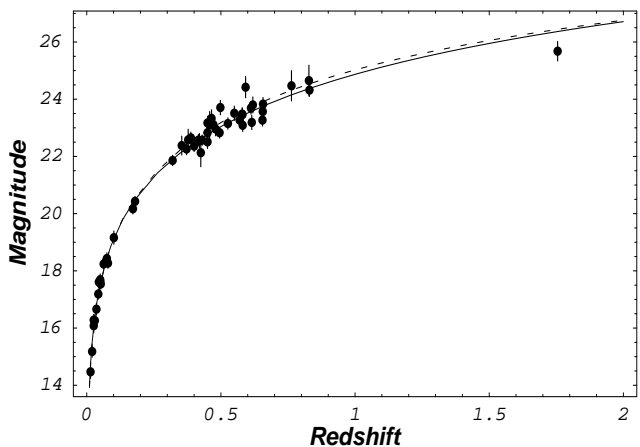,height=2.3in,width=3.3in,angle=0}
{\caption{\small The Hubble diagram for 55 supernovae: the theoretical
curves represent the best fitting model for the case $k=0$ obtained
by fixing $\Omega_{\rm m0}$ to 0.33 (solid curve)
and the favoured standard model $\Omega_{\rm m0}=1-\Omega_{\Lambda 0}=0.34$
(dashed curve). The empty circles represent the six excluded supernovae
by Perlmutter et al. in their primary fit C \cite{perl}.}}
\end{figure}

We further note that the estimate of the parameter
${\cal M}$, which is usually referred to as the {\it Hubble
constant-free-absolute magnitude}, seems to be model independent, unlike
the other parameters ($\Omega_i$) which very much depend on the models
considered. For instance,
the SN dataset, which we are considering here, has been fitted to a variety
of models $-$ the standard FRW models \cite{vishwa2},
kinematical $\Lambda$-models \cite{vishwa2}, quintessence models
\cite{vishwa1}, quasi-steady state model \cite{jvn}, etc. All give
the same ${\cal M}\approx 24$, for low as well as high redshift
supernovae \cite{vishwa3}.
This is remarkable and consistent with the primary assumption that the
supernovae Ia are {\it standard candles}.
To fix ideas, let us note that ${\cal M}=24$ gives the corresponding
absolute magnitude $M$ at peak luminosity of the supernovae as $M=-19.3 $
for $H_0 = 65$  km s$^{-1}$ Mpc$^{-1}$ and $M=-19.1 $
for $H_0 = 72$  km s$^{-1}$ Mpc$^{-1}$, which are in the right region
for SN peak magnitudes \cite{saha}.

Summarizing, the modification caused by the inclusion of the brane
scalar curvature in the action amounts to defining an effective dynamical
curvature which is different from the geometric curvature.
This essentially shifts the curvature parameter $k$ to the
\emph{effective curvature parameter} $(k - \alpha)$ in the
Friedmann equation,
hence making the difference even at the present epoch, which can be
tested against the present observations. It seems to come out very well.
This provides a welcome leverage  which could be
maneuvered for having a  comfortable age for the universe. 
Our ongoing investigations indicate that the model
seems to explain the CMBR anisotropy measurements and radio sources data
successfully \cite{next}.

Although the degeneracy in the parameter space is large, which is due
to inadequacy of the present data, the situation will improve when more
precise data with more points at $z > 1$ become available. It would then
narrow down the degeneracy pinning the parameters more accurately.
 This might be accomplished by the proposed SNAP
({\it SuperNova Acceleration Probe}) project.

\emph{Acknowledgements:} We thank J. V. Narlikar and R. Maartens for
useful comments and discussions. PS thanks CSIR for a
research fellowship and RGV thanks DAE for his Homi Bhabha postdoctoral fellowship.

\end{document}